# Towards a framework for evaluating scientific disruption indices

Tao Zhou[1], Chi Zhang[1], Yilin Bi

CompleX Lab, School of Computer Science and Engineering, University of Electronic Science and Technology of China, Chengdu 611731, P. R. China

**Abstract**: The disruption index *D* uses subsequent citation patterns to quantify the disruptiveness of a focal paper relative to the prior work it builds on. It has become an important research tool in the science of science. Its validity and applicability remain the subject of ongoing debate, and numerous variants have been proposed. The recent introduction of the *1D* index has intensified this debate. We argue that the most pressing need is a transparent, fair, and practical framework for evaluating disruption indices, including *D* and *1D*. Accordingly, we propose a framework for evaluating disruption indices based on two key criteria: how well an index identifies papers independently verified to be highly disruptive within a large pool of candidate papers, and whether it improves identification accuracy beyond what can be achieved using total citation counts alone. We also consider three secondary criteria: whether an index is associated with paper characteristics that are not expected to reflect disruptiveness, such as reference list length, whether it distinguishes original research articles from review articles, and whether it is temporally stable. Our empirical results show that several widely studied disruption indices each have substantial limitations. None of those indices satisfies both key criteria. We conclude by outlining the limitations of our framework and discussing what is needed to develop a more comprehensive one.

## Introduction

Scientific progress includes both incremental refinement within prevailing paradigms and rare episodes of disruption, which break the existing theoretical tradition and reconstruct the research field from new foundations [1]. Funk and Owen-Smith [2] developed an index for differentiating consolidating and disruptive patent technologies. Shortly afterwards, Wu, Wang and Evans [3] adopted this index to quantify disruption in scientific publications. This index, termed the disruption index (*D*), relies solely on subsequent citations. Specifically, for a focal paper, the disruption index is defined as

$$D=\frac{N_i-N_j}{N_i+N_j+N_k}, \quad (1)$$

where $N_i$ is the number of subsequent articles that cite the focal paper but none of its references, $N_j$ counts subsequent articles that cite both the focal paper and at least one of its references, and $N_k$ counts subsequent articles that cite at least one of the focal paper's references but not the focal paper. It is both intuitive and computationally tractable, and has therefore rapidly become the most popular tool to analyze the disruptiveness of scientific publications [4-8]. Meanwhile, debates over the validity and adaptability of the disruption index persist [9-11], accompanied by the proposal of a set of its variants [12-16].

[1] These two authors contribute equally to this work.

Recently, the paper by Huang and Xie [17] has intensified this ongoing debate. Huang and Xie argued that the *D* index declines mechanically as reference lists lengthen. Therefore, the drop in disruptiveness of large teams, as claimed by Wu, Wang and Evans [3], arises simply because large teams cite more references. In addition, Huang and Xie proposed a new metric named *1D*. The basic idea is to compute the pairwise *D* index between the focal paper and each of its cited references, and then take the average. Huang and Xie further found that the *1D* index achieves far higher accuracy than the *D* index for the task of identifying papers with widely recognized disruptive innovations, such as those awarded the Nobel Prize and Turing Award. Lin, Yang and Wu [18] soon drafted a counterargument paper, pointing out that the *1D* index is also strongly correlated with reference-list length, albeit in the opposite direction. Moreover, the *1D* index assigns equal weight to each cited reference when computing the average. This makes it easy for positive scores from numerous low-cited references to overwhelm the negative scores from a small set of key prior works, thereby classifying incremental knowledge-building contributions as disruptive.

The most effective way to resolve such debates is to develop a transparent, fair, and practical framework for evaluating disruption indices. We propose an evaluation framework consisting of two critical criteria (gold standards) and three secondary yet important criteria (silver standards). We test several well-known disruption indices under this framework and find that none satisfy both gold standards simultaneously. While this is only a preliminary framework, the results strongly suggest that we need to reconsider how to accurately characterize the disruptiveness of a scientific paper. If this issue cannot be adequately resolved, existing findings and future results derived from current indices regarding disruption in scientific research are questionable. We conclude by discussing how to construct a more refined framework and how to design a more accurate index.

## Framework

We denote the disruption index to be evaluated as $D^*$. If $D^*$ is valid, it should at least be able to identify widely recognized highly disruptive papers. Let $A$ be the set of all papers to be analyzed, and $A_d$ the set of highly disruptive papers that have been identified. $A_d$ may be identified via text analysis, expert annotation, surveys of scholars, major awards, or even a collaboration between large language models and experts. Critically, the information used to compute $D^*$ must not be reused for constructing $A_d$; otherwise, the argument becomes circular. Note that we do not require every paper in $A_d$ to be more disruptive than every paper in $A \backslash A_d$. It is sufficient that the disruption level of $A_d$ is significantly higher than the average level over $A \backslash A_d$.

Provided $A_d$, we can use the area under the receiver operating characteristic curve (AUC) [19] to quantify the ability of $D^*$ in identifying highly disruptive papers. In this context, the AUC can be interpreted as the probability that the $D^*$ value of a randomly selected paper from $A_d$ is greater than the $D^*$ value of a randomly selected paper from $A \backslash A_d$ [20]. AUC will approach 0.5 if the index $D^*$ gives completely random values to papers, and thus to what extent AUC exceeds 0.5 indicates how much better $D^*$ performs than random guessing. Even under the most minimal requirement, a valid disruption index should achieve higher accuracy than random guessing in identifying highly disruptive papers. This yields the first gold standard.

**GS1**: If a subset with significantly higher disruption has been identified from the population of papers via credible external means, treating this subset as positive samples, the AUC of the disruption index for identifying these positive samples should be greater than 0.5.

Since total citations are the primary and most intuitive metric used by the academic community to evaluate a paper, a valuable disruption index, when incorporated into the task of identifying highly disruptive papers, should significantly improve the accuracy achieved using total citations alone. This yields the second gold standard.

**GS2**: The disruption index can significantly improve the accuracy in identifying highly disruptive papers at the presence of total citations.

This paper adopts an intuitive and concise approach, in which the disruption index $D^*$ to be evaluated and total citations $C$ are linearly combined to form a new indicator, as

$$D_C = a\widetilde{D^*} + (1-a)\widetilde{C}, \tag{2}$$

where $a \in [0,1]$ is a tunable parameter, and $\widetilde{D^*}$ and $\widetilde{C}$ denotes the values after min-max normalization, for example,

$$\widetilde{C} = \frac{C - C_{min}}{C_{max} - C_{min}}, \tag{3}$$

where $C_{max}$ and $C_{min}$ denotes the maximal and minimal total citations across all considered papers.

Denote by $\Delta AUC$ the improvement resulted from the introduction of the disruption index $D^*$, namely

$$\Delta AUC = AUC(C, D^*) - AUC(C), \tag{4}$$

with $AUC(C)$ and $AUC(C,D^*)$ being the accuracy in identifying papers in $A_d$ using only total citations and using the linear combination of total citations and the index $D^*$. We expect that the introduction of $D^*$ can lead to a substantial increase in AUC, which is expressed as

$$\Delta AUC \geq pAUC(C). \quad (5)$$

This work considers a relatively lenient requirement: the introduction of $D^*$ produces an AUC increment of no less than 1%, namely $p$=0.01.

We next consider three secondary yet important criteria (silver standards).

**SS1**: The disruption index is not correlated with paper characteristics that are not expected to reflect disruptiveness.

This paper will examine the association between the disruption index and reference list length, since this attribute is supposed to be unrelated to a paper's disruptiveness. Notice that, Huang and Xie's criticism of the $D$ index initially targets its negative correlation with reference list length.

Since review articles are inherently designed to synthesize and consolidate prior knowledge and, in most cases, do not report novel research findings, their disruptiveness should be lower than that of original research articles. Based on this reasoning, we propose the following criterion.

**SS2**: On average, the disruption index should assign higher scores to original research articles than to review articles.

Disruptiveness is an intrinsic property of a paper, which is fixed once the paper is published and should not change over time. Since disruption indices rely on citation data from subsequent papers, our estimate of a paper's disruptiveness may evolve as time passes. Nevertheless, the index should gradually stabilize after sufficient data accumulates. Given a disruption index $D^*$, if we define $AUC(t_1,t_2)$ ($t_1<t_2$) as the probability that the value of $D^*$ calculated from citation data accumulated up to $t_2$ years after the publication of the focal paper is higher than that calculated using data up to $t_1$ years after the publication, say

$$AUC(t_1,t_2) = \frac{1}{N(t_2)}\sum_{i=1}^{N(t_2)} I(D^*(p_i,t_2) > D^*(p_i,t_1)) + 0.5I(D^*(p_i,t_2) = D^*(p_i,t_1)), \quad (6)$$

where $N(t)$ is the number of papers having published for at least $t$ years, $i$ runs over these papers, $D^*(p_i,t)$ is the value of $D^*$ for the $i$th paper and its citation records up to $t$ years after the publication, and $I(q)$ is the indicator function, which equals 1 if condition $q$ holds and 0 otherwise. If $D^*$ exhibits an upward trend, $AUC(t_1,t_2)$ will be larger than 0.5, while if $D^*$ exhibits a downward trend, $AUC(t_1,t_2)$ will be smaller than 0.5. In contrast, a temporally stable index should produce a value of $AUC(t_1,t_2)$ close to 0.5. Accordingly, we propose the final criterion.

**SS3**: The disruption index should be temporally stable, with neither an upward nor a downward trend over time.

## Results

In addition to $D$ and $1D$, we consider six well-studied disruption indices [3,12,13]. The definitions of the eight indices are presented in Table 1. We analyze publication and citation records from 1990 to 2020 in the data set called Microsoft Academic Graph (MAG) [21]. Only papers receiving at least one citation are included in the following analysis, yielding a filtered data set of 10,731,740 papers. Motivated by Huang and Xie [17], we assume that the papers associated with highly prestigious scientific prizes are more disruptive and thus an effective disruptive index could identify them from a large corpus of papers.

**Table 1**: Definitions and explanations of the eight disruption indices under consideration.

| Index | Definition and Explanation |
| --- | --- |
| $D=\frac{N_i-N_j}{N_i+N_j+N_k}$ | $N_x=\|\{p\|p\in \text{category } x\}\|$. Category $i$ includes subsequent articles that cite the focal paper (FP) but none of its references. Category $j$ includes subsequent articles that cite both the FP and at least one of its references. Category $k$ includes subsequent articles that cite at least one of the FP's references but not the FP. |
| $D_5=\frac{N_i-N_j^5}{N_i+N_j^5+N_k}$ | $N_j^5=\|\{p\|p \text{ cites the FP and at least five of the FP's references}\}\|$. |
| $D^{nk}=\frac{N_i-N_j}{N_i+N_j}$ | $N_k$ is excluded from the denominator. |
| $D_5^{nk}=\frac{N_i-N_j^5}{N_i+N_j^5}$ | $N_j^5$ is defined as in $D_5$ and $N_k$ is excluded from the denominator. |
| $D^{ns}=\frac{N_i^{ns}-N_j^{ns}}{N_i^{ns}+N_j^{ns}+N_k^{ns}}$ | $N_x^{ns}=\|\{p\|p\in \text{category } x, \text{ and } p \text{ is not a self-citing article}\}\|$, with $x\in\{i,j,k\}$. |
| $D^{link}=\frac{L_i-L_j}{L_i+L_j+L_k}$ | $L_x$ counts citation links from the articles in category $x$ to FP or any of its references, with $x\in\{i,j,k\}$. |
| $ND=\frac{N_i}{N_i+N_j+\widehat{N_k}}$ | $\widehat{N_k}=\frac{\mathrm{N_k}}{\|\mathrm{Ref}_{FP}\|}$ and $\|\mathrm{Ref}_{FP}\|$ denotes the number of FP's references. |
| $1D=\frac{1}{\|\mathrm{Ref_{FP}}\|}\sum_{r\in \mathrm{Ref_{FP}}}\frac{N_i^{(r)}-N_j^{(r)}}{N_i^{(r)}+N_j^{(r)}+N_k^{(r)}}$ | $1D$ evaluates the FP's disruptiveness relative to each reference separately and takes the average of these scores, giving all references equal weight. $N_x^{(r)}=\|\{p\|p\in \text{category } x, \text{ and only citations related to reference } r \text{ are considered}\}\|$, with $x\in\{i,j,k\}$. |

Among these 10,731,740 papers, 35 are associated with Turing Award, 196 with Lasker Awards, 63 with Nobel Prize, 51 are designated as Physical Review Milestones (PR Milestones for short), and 247 are featured as Science Breakthroughs of The Year (Breakthroughts for short). We treat these five categories of honored papers as positive examples $A_d$, and the full set of 10,731,740 papers under study as set $A$. Based on this setup, we compute the AUC values for all considered disruption indices and total citation count $C$. As shown in Table 2, the total citation count achieves the highest AUC value in every category. Among the eight disruption indices considered in our work, $D_5$, $ND$ and $1D$ substantially outperform the other five indices. These are also the only three indices whose AUC values are greater than 0.5 across all categories. In other words, only these three indices can pass the gold standard GS1.

**Table 2**: The AUC values for the five categories of honored papers, respectively obtained by the eight disruption indices and the total citation count. The last row show the average values. The bold-faced values correspond to the best-performing disruption index.

| $A_d$ | $D$ | $D_5$ | $D^{nk}$ | $D_5^{nk}$ | $D^{ns}$ | $D^{link}$ | $ND$ | $1D$ | $C$ |
|---|---|---|---|---|---|---|---|---|---|
| Turing Award | 0.6317 | **0.7484** | 0.6703 | 0.6261 | 0.6852 | 0.5634 | 0.7346 | 0.7253 | 0.8250 |
| Lasker Awards | 0.3051 | 0.7516 | 0.4790 | 0.3864 | 0.4862 | 0.1887 | 0.8166 | **0.9003** | 0.9691 |
| Nobel Prize | 0.5320 | 0.8675 | 0.6305 | 0.4997 | 0.6441 | 0.2440 | 0.9116 | **0.9770** | 0.9986 |
| PR Milestone | 0.6213 | **0.8895** | 0.6534 | 0.5830 | 0.6798 | 0.5810 | 0.8862 | 0.8764 | 0.9852 |
| Breakthroughs | 0.2482 | 0.8046 | 0.4453 | 0.4135 | 0.4292 | 0.1287 | 0.8200 | **0.9108** | 0.9868 |
| Average | 0.4677 | 0.8123 | 0.5757 | 0.5018 | 0.5849 | 0.3411 | 0.8338 | 0.8780 | 0.9529 |

**Table 3**: The AUC increment for identifying honored papers achieved by linearly combining the total citation count $C$ with each disruption index individually, relative to the AUC using only $C$.

| $\Delta AUC$ | $D$ | $D_5$ | $D^{nk}$ | $D_5^{nk}$ | $D^{ns}$ | $D^{link}$ | $ND$ | $1D$ |
|---|---|---|---|---|---|---|---|---|
| Turing Award | 0.0246 | 0.0044 | 0.0340 | 0.0625 | 0.0231 | 0.0346 | 0.0013 | 0 |
| Lasker Awards | 0 | 0 | 0 | 0 | 0 | 0.0002 | 0 | 0 |
| Nobel Prize | 0 | 0 | 0 | 0 | 0 | 0 | 0 | 0 |
| PR Milestone | 0.0002 | 0.0016 | 0 | 0 | 0.0014 | 0 | 0.0024 | 0.0046 |
| Breakthroughs | 0 | 0 | 0 | 0 | 0 | 0 | 0 | 0 |

Table 3 shows $\Delta AUC$ for each of the eight disruption indices. In most cases, adding a disruption index following Eq. (2) yields no improvement in the accuracy of identifying honored papers. Where improvements do occur, they are marginal. Under the criterion defined by Eq. (5), $D$, $D^{nk}$, $D_5^{nk}$, $D^{ns}$, and $D^{link}$ satisfy GS2 only for papers associated with the Turing Award. Although the $1D$ index overall achieves higher standalone identification accuracy than the other seven disruption indices, it delivers no further accuracy gains conditional on known total citations. These observations may arise either because disruption indices provide no additional information gain over total citations for identifying honored papers, or because total citations alone already enable highly accurate identification. One critical caveat is that disruption indices are designed to detect highly disruptive papers, rather than papers associated with the Turing Award or the Nobel Prize. Accordingly, the definition of $A_d$ adopted in this work, which follows Reference [17], constitutes only an expedient choice at the initial stage of related research. We further discuss data-set-related issues in greater depth in the final section.

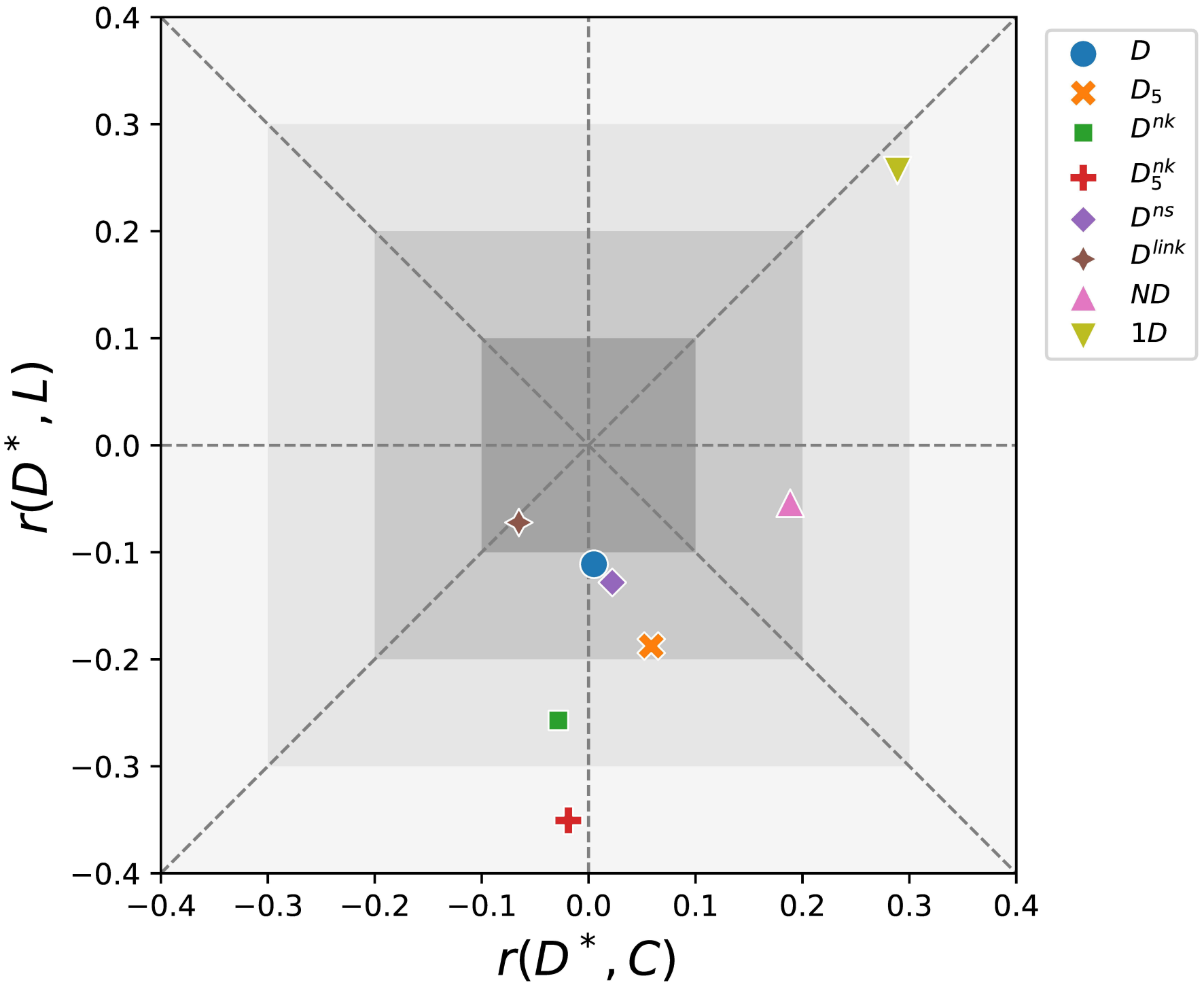

**Figure 1**: Pearson correlations of the eight disruption indices with total citation count (x-axis) and with reference list length (y-axis).

Next, we examine the association between the values of disruption indices and features of target papers that are unrelated to disruptiveness. Since the relationship between disruption indices and reference list length is the focus of the recent debates [17,18], we use reference list length as an illustrative example. For a given disruption index $D^*$ under analysis, we compute the Pearson correlation $r(D^*,L)$ between its values and the reference list length $L$ across the 10,731,740 papers. Since GS2 concerns the information gain of $D^*$ relative to $C$, we additionally calculate the Pearson correlation $r(D^*,C)$ between $D^*$ and $C$. As shown in Figure 1, only the absolute values of $r(D^{link}, L)$ and $r(ND, L)$ are less than 0.1. Although Huang and Xie criticized that the $D$ index has a significant correlation with reference list length [17], the association between *1D* and reference list length is even stronger, albeit in the opposite direction: $D$ is negatively correlated with $L$, whereas *1D* is positively correlated with $L$. In fact, among the eight disruption indices analyzed in this work, *1D* also exhibits the strongest correlation with total citations. This partly explains why *1D* alone can identify honored papers with relatively high accuracy, yet adding *1D* conditional on known total citations yields no further improvement in identification performance. Taking $r < 0.1$ as the threshold, only $D^{link}$ and *ND* satisfy SS1.

Review articles synthesize and consolidate existing research and are therefore expected to exhibit low disruption. Without article type labels in the MAG dataset, we distinguish review articles from original research articles using a natural language processing (NLP) model (SciBERT) based on their titles and abstracts [22]. The fine-tuned SciBERT model achieves a precision of 90.45% and an F1-score of 92.79%. We apply the model to our analytical sample, and identify 8,939,083 original research articles and 1, 792,657 review articles. As shown in Figure 2, six out of the eight disruption indices assign higher average values to original research articles. By contrast, *ND* and *1D* indicate that review articles are, on average, more disruptive. In other words, *ND* and *1D* do not satisfy SS2.

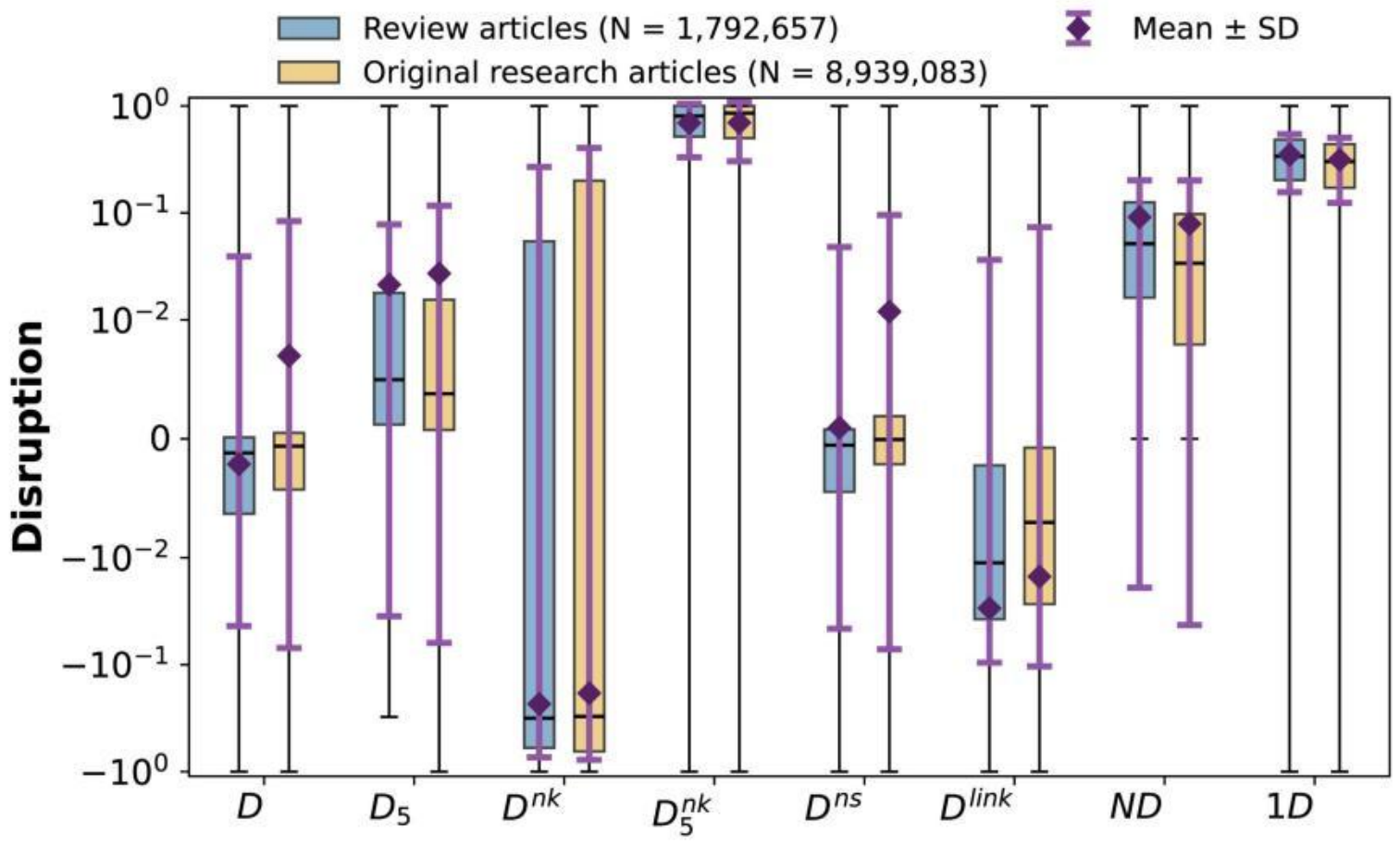

| | Review articles | Original research articles |
|---|---|---|
| $D$ | -0.002 | 0.007 |
| $D_5$ | 0.021 | 0.027 |
| $D^{nk}$ | -0.233 | -0.185 |
| $D_5^{nk}$ | 0.692 | 0.698 |
| $D^{ns}$ | 0.001 | 0.012 |
| $D^{link}$ | -0.030 | -0.015 |
| $ND$ | 0.091 | 0.079 |
| $1D$ | 0.351 | 0.314 |

**Figure 2**: Distributions of the eight disruption indices across review articles and original research articles. The total sample consists of 10,731,740 papers, including 1,792,657 review articles (16.7%, blue) and 8,939,083 original research articles (83.3%, yellow). The horizontal axis lists the eight disruption indices in sequence, and the vertical axis shows the values of each index on a logarithmic scale. The upper and lower bounds of each box denote the 75th and 25th percentiles, the black horizontal line inside each box marks the median, and the endpoints of the black whiskers correspond to the maximum and minimum values of each group. Purple diamonds represent the mean values, and the purple error bars indicate one standard deviation from the mean. The table on the right directly reports the mean values of the eight indices for review articles and original research articles.

Figure 3 shows how the mean values of the eight disruption indices change with *t*, using only citation data accumulated up to *t* years after the publication of each focal paper. Visually, $D$, $D^{nk}$, $D_5^{nk}$, $D^{ns}$, and $D^{link}$ exhibit an upward trend over time, while $D_5$, *ND*, and *1D* show no clear trend. None of the curves display a downward trend as time increases. To obtain more quantitative results, we compare the scenarios at $t = 5$ and $t = 10$ and compute AUC(5,10) as defined in Eq. (6). The sample comprises 5,984,118 papers in the MAG dataset that have been published for at least 10 years (papers published in 2010 or earlier), i.e., $N(10) = 5{,}984{,}118$. As shown in Figure 3, the AUC(5,10) values for $D$, $D^{nk}$, $D_5^{nk}$, $D^{ns}$, and $D^{link}$ are all greater than 0.6, meaning that, on average, for more than 60% of papers, their disruption value at 10 years post-publication exceeds their disruption value at 5 years post-publication. This clearly violates the expected property of temporal stability. Accordingly, only $D_5$, *ND*, and *1D* satisfy SS3.

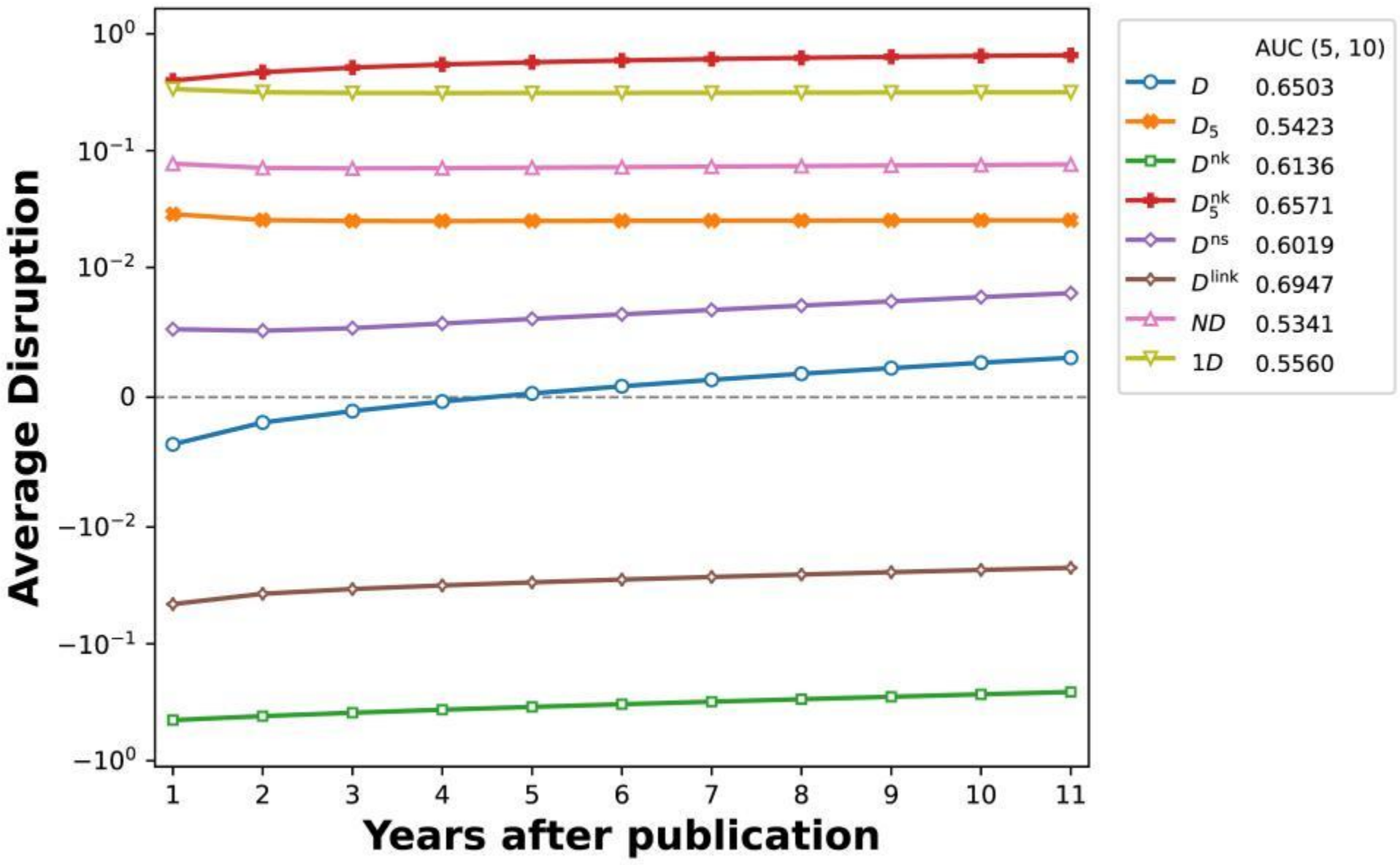

**Figure 3**: The temporal trend of average disruption values calculated from citation data accumulated up to $t$ years after the publication of each focal paper. The eight curves correspond to the results of the eight disruption indices. The upper-right panel presents the values of $AUC(5,10)$.

**Table 4**: Summary of whether the eight disruption indices pass the five criteria.

| | GS1 | GS2 | SS1 | SS2 | SS3 |
|---|---|---|---|---|---|
| $D$ | | ✓ | | ✓ | |
| $D_5$ | ✓ | | | ✓ | ✓ |
| $D^{nk}$ | | ✓ | | ✓ | |
| $D_5^{nk}$ | | ✓ | | ✓ | |
| $D^{ns}$ | | ✓ | | ✓ | |
| $D^{link}$ | | ✓ | ✓ | ✓ | |
| $ND$ | ✓ | | ✓ | | ✓ |
| $1D$ | ✓ | | | | ✓ |

## Discussion

Table 4 summarizes the performance of the eight indices under the five criteria. It can be observed that none of the indices pass all five criteria, and not a single index satisfies both gold standards simultaneously. It appears that different indices exhibit distinct properties and limitations. Moreover, D and 1D do not demonstrate performance that is markedly better than each other or the remaining indices. Accordingly, it remains an open challenge to design disruption indices for more accurate detection of highly disruptive papers while providing information gain beyond total citations. It is possible that combinations of $N_i$, $N_j$, $N_k$ or their simple variants alone cannot adequately address this problem. At present, at least two promising research avenues exist. The first is to define disruption indices based on the position of the focal paper within the full citation network, rather than relying merely on simple counts [14,16]. The second leverages the text of the focal paper to improve the accuracy of disruption evaluation. Naturally, both approaches inevitably sacrifice some parsimony and elegance [23,24]. Alternatively, we may also attempt to design disruption indices by modeling how researchers search for, read, and cite the references and citing works of a focal paper after reading it.

Based on the empirical results of this study, we observe several intriguing patterns. For example, all indices that satisfy GS2 show an upward trend over time. As another example, the eight indices can be divided into two groups: the five indices ($D$, $D^{nk}$, $D_5^{nk}$, $D^{ns}$, and $D^{link}$) that pass GS2, and the three indices ($D_5$, $ND$, and $1D$) that pass GS1. Indices within the same group exhibit more similar properties. Are these patterns merely coincidental, or do they stem from unknown mechanisms? If additional indices are incorporated, what will the correlation structure among these indices look like? Will they also form clusters? These questions require further large-scale empirical analyses (referring to analogous studies on node centrality metrics [25]).

Although our goal is to develop a relatively comprehensive framework for evaluating disruption indices to support future research in this field, this work represents only a preliminary attempt, and the proposed framework still has evident limitations. For instance, we have not constructed a dedicated dataset for assessing disruption indices. Furthermore, the combination scheme (linear combination) of disruption indices and total citation count used to quantify information gain is merely the most intuitive and simplest among many plausible alternatives. In addition, the external factor we examined (reference list length) is only one of numerous potential confounding factors.

Among these issues, the most critical and urgent one is to build a better dataset for evaluating disruption indices. Specifically, we need to construct larger sets of $A_d$ consisting of papers that are statistically characterized by substantially above-average disruptiveness (these sets may be constructed separately for different research fields).

The current honored-paper sets, which comprise only tens to hundreds of papers, are poorly suited for our purpose. First, disruption indices quantify the disruptive level of a paper, rather than the probability that the paper constitutes the major contribution underlying a Nobel Prize. The analogy is that we cannot expect a general physical-fitness index for middle-school students to directly identify prospective world champions. Second, honored papers are inherently highly influential within their fields and accrue enormous citation counts. As a result, total citation count already serves as a naturally strong predictor, making it extraordinarily challenging to design an index that delivers additional information gain conditional on total citations. Accordingly, our top priority is to assemble several collections of papers that are highly disruptive yet do not rise to the level of work that underpins a Nobel Prize. Manual annotation, textual analysis, and a credible large-language-model jury with disagreement arbitration mechanisms may all assist in this endeavor.